# WYKORZYSTANIE REKONFIGUROWALNYCH INTELIGENTNYCH MATRYC ANTENOWYCH W ŁĄCZU DOSYŁOWYM SIECI 5G/6G WYKORZYSTUJĄCEJ BEZZAŁOGOWE STATKI POWIETRZNE

## ON THE USE OF RECONFIGURABLE INTELLIGENT SURFACES FOR BACKHAULING IN UAV-ASSISTED 5G/6G NETWORK

Salim Janji[1]; Paweł Sroka[2]; Adrian Kliks[3]

[1] Politechnika Poznańska, Intytut Radiokomunikacji, Poznań, salim.janji@doctorate.put.poznan.pl
[2] Politechnika Poznańska, Intytut Radiokomunikacji, Poznań, pawel.sroka@put.poznan.pl
[3] Politechnika Poznańska, Intytut Radiokomunikacji, Poznań, adrian.kliks@put.poznan.pl

**Streszczenie**: Drony, dzięki możliwości ich szybkiego rozmieszczenia w trudnym terenie, uważane są za jeden z kluczowych elementów systemów bezprzewodowych 6G. Jednak w celu wykorzystania ich jako punkty dostępowe sieci konieczne jest zapewnienie łącza dosyłowego o odpowiedniej przepustowości. Dlatego w niniejszym artykule rozważane jest zwiększenie zasięgu sieci bezprzewodowej przez zapewnienie łącza dosyłowego dla końcowego punktu dostępowego z wykorzystaniem określonej liczby dronów-przekaźników oraz rekonfigurowalnych inteligentnych matryc antenowych (RIS). Zaprezentowane wyniki badań symulacyjnych pokazują, że użycie RIS pozwala na znaczące zwiększenie zasięgu sieci bez konieczności stosowania dodatkowych przekaźników.

**Abstract**: Unmanned Aerial Vehicles, due to the possibility of their fast deployment, are considered an essential element of the future wireless 6G communication systems. However, an essential enabler for their use as access points is to provide a sufficient throughput wireless backhaul link. Thus, in this paper we consider the aspect of extension of network coverage with the use of drone-based relaying and reconfigurable intelligent surfaces (RIS) for backhauling. Presented results of simulation experiments indicate that the use of RIS allows for significant improvement of network coverage without the need to use additional relays.

**Słowa kluczowe**: łącze dosyłowe, przekaźniki, drony, RIS, 6G

**Keywords**: backhaul, relays, RIS, UAV, 6G

## 1. WSTĘP

Jednym z podstawowych zastosowań systemów bezprzewodowych piątej generacji (5G) jest zapewnienie usług szerokopasmowej transmisji danych (ang. *enhanced mobile broadband* – eMBB). Rozwiązaniem pozwalającym na realizację tego zadania jest rozmieszczenie bardzo dużej liczby stacji bazowych (ang. *base station* – BS) i punktów dostępowych (ang. *access point* – AP), tworząc w ten sposób sieć o bardzo dużym zagęszczeniu (ang. *ultra-dense network* – UDN). Jednak użycie tak dużej liczby BS i AP wymaga zapewnienia odpowiedniej jakości łączy dosyłowych w celu wymiany danych z siecią rdzeniową [1], co może okazać się niemożliwe lub nieopłacalne w przypadku stosowania łączy przewodowych.

Rozwiązaniem problemu zapewnienia łącza dosyłowego w UDN może być użycie komunikacji bezprzewodowej oraz bezzałogowych statków powietrznych (ang. *unmanned aerial vehicle* –UAV), czyli dronów, działających jako przekaźniki radiowe [2]. Rozmieszczenie UAV w sposób pół-statyczny pozwala utworzyć sieć tymczasowych łączy dosyłowych między stacją bazową makro i AP tworzącymi małe komórki. Dzięki odpowiedniemu zlokalizowaniu dronów-przekaźników (ang. *drone relay station* – DRS) możliwe jest uzyskanie łączy z bezpośrednią widocznością (ang. *line-of-sight* –LoS), co zapewnia większą przepustowość niż w przypadku łączy bez bezpośredniej widoczności.

Zagadnienie rozmieszczenia DRS i sterowania bezprzewodowym łączem dosyłowym z wieloma przeskokami jest skomplikowane, gdyż wpływa na nie szereg ograniczeń, takich jak: rozmiar UAV, jego mobilność, zużycie energii czy wolumen przesyłanych danych. Dodatkowo, w środowisku miejskim nie zawsze możliwe jest zapewnienie łączności typu LoS z ograniczoną liczbą dronów. Dlatego jednym z rozwiązań wspomagających użycie UAV może być zastosowanie rekonfigurowalnych inteligentnych matryc antenowych (ang. *reconfigurable inteligent surfaces* – RIS), pozwalających na stworzenie wirtualnych połączeń LoS między nadajnikiem i odbiornikiem, dla których w normalnych warunkach bezpośrednia widoczność nie występuje [3]. RIS to matryca zawierająca dużą liczbę elementów, których zadaniem – w najprostszym ujęciu - jest odbicie docierającego sygnału radiowego, jednocześnie zmieniając jego amplitudę, częstotliwość lub fazę [4]. Dzięki użyciu RIS pojawia się możliwość oddziaływania na propagację sygnału zwiększając w ten sposób przepustowość, niezawodność czy zasięg transmisji. Zaletą RIS jest łatwość ich rozmieszczenia poprzez montaż na ścianach budynków, bilbordach, pojazdach czy statkach powietrznych. Dodatkowo są one efektywniejsze energetycznie od tradycyjnych przekaźników i w pełni kompatybilne z istniejącymi systemami komunikacji bezprzewodowej. W przypadku wykorzystania RIS w komunikacji z UAV, pozwalają one na utworzenie krótszych, niezawodnych wirtualnych łączy LoS, w szczególności w środowisku miejskim. Łączna optymalizacja położenia UAV i alokacja zasobów radiowych uwzględniająca użycie RIS pozwala też na minimalizację zużycia energii w dronach. Zagadnienie wykorzystania RIS w realizacji łącza dosyłowego z wykorzystaniem dronów jest stosunkowo nowym rozwiązaniem, w związku z czym dostępne są jedynie nieliczne publikacje. Rozwiązanie zaprezentowane w [5] wykorzystuje matryce umieszczone na statkach powietrznych dużych wysokości, optymalizując łącze dosyłowe w zakresie efektywności energetycznej przez podział matrycy i odpowiednie fazowanie jej elementów. Z kolei w [6] rozważono RIS zamontowane na dronach, stosując model wielorękiego bandyty do optymalizacji łącza dosyłowego w paśmie fal milimetrowych.

W niniejszej pracy, rozwijając badania przedstawione w [7], rozważany jest problem rozmieszczenia DRS i wyboru ścieżki transmisji od stacji bazowej makro do AP w celu realizacji łącza dosyłowego w środowisku miejskim, gdzie drony znajdują się poniżej wysokości budynków. Realizacja łącza odbywa się z wieloma przeskokami z wykorzystaniem pasma fal milimetrowych i określonej minimalnej przepustowości. Zakładane jest też wykorzystanie RIS rozmieszczonych na ścianach wybranych budynków, co pozwala na zwiększenie zasięgu sieci przy użyciu stałej liczby DRS.

Dalsza część artykułu przedstawia się następująco: w rozdziale 2. przedstawiono model rozważanego systemu oraz sformułowano problem optymalizacyjny. W rozdziale 3 przedstawiono proponowane rozwiązanie oraz wyniki badań symulacyjnych jego zastosowania. Rozdział 4. podsumowuje wyniki dotychczasowych badań.

## 2. MODEL SYSTEMU I PROBLEM OPTYMALIZACYJNY

### 2.1. Model systemu

W niniejszej pracy rozważane jest zagadnienie wykorzystania UAV dla realizacji bezprzewodowego łącza dosyłowego z wieloma przeskokami w środowisku miejskim, gdzie dostępna jest tylko jedna stacja bazowa makro (MBS) odpowiedzialna za łączność z siecią rdzeniową. W związku z tym poszerzenie zasięgu sieci realizowane jest z wykorzystaniem mobilnych punktów dostępowych (AP) w postaci dronów połączonych bezprzewodowo z MBS. Zakładamy dostępność pasma o szerokości B w zakresie częstotliwości fal milimetrowych w celu realizacji łącza dosyłowego, a także wykorzystanie maksymalnie N dronów, które mogą działać jako AP lub DRS. Rozważany obszar miejski został zaprojektowany na podstawie analizy fragmentu zabudowy miasta Poznań pobranego z serwisu OpenStreetMap, co przedstawiono na rys. 1. Drony, po określeniu ich oczekiwanej lokalizacji, działają w sposób pół statyczny – nie zmieniają swojego położenia w trakcie realizacji transmisji. Dodatkowo, w systemie możliwe jest zastosowanie do R matryc RIS umieszczonych na wybranych ścianach budynków w celu poprawy zasięgu pojedynczego przeskoku łącza dosyłowego.

Ze względu na wymagania minimalnej przepustowości tworzonego łącza ($C_{min}$) zakłada się, że łącze dosyłowe dla wybranej lokalizacji AP jest dostępne jeśli dla każdego z przeskoków, tj. dla każdego łącza między MBS i DRS, dwoma DRS i między DRS i AP, można osiągnąć założoną minimalną przepustowość. Dodatkowo ze względu na wykorzystanie częstotliwości fal milimetrowych założono, że straty wynikające z odbić fal od obiektów innych niż RIS są na tyle wysokie, że nie pozwalają na realizację łącza dosyłowego, w związku z czym wymagane jest zawsze łącze LoS. Natomiast możliwa jest

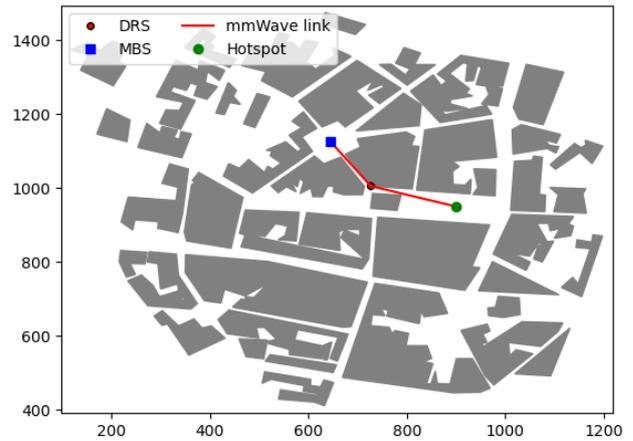

*Rys. 1. Ilustracja rozważanego scenariusza - schemat układu budynków, stacji bazowej makro (MBS) i rozważanej lokalizacji DRS i AP (hotspot) – opracowany na podstawie struktury zabudowy miasta Poznań*

transmisja w konfiguracji wirtualnego LoS, tzn. wykorzystując odbicie sygnału od RIS - w tym przypadku całkowite tłumienie takiego łącza jest sumarycznym tłumieniem dwóch ścieżek: od nadajnika do RIS i od RIS do odbiornika. W związku z tym tłumienie pojedynczego przeskoku łącza może być wyznaczone następująco:

- w przypadku łącza bezpośredniego LoS (bez RIS) [8]:

$$PL_{direct}^{dB}(d) = PL(d_0) + 10\alpha \cdot \log(d) \quad (1)$$

gdzie d to odległość między nadajnikiem i odbiornikiem, $PL(d_0)$ to tłumienie wolnej przestrzeni dla $d_0 = 5\ m$, a $\alpha$ to współczynnik tłumienia ścieżki (wykładnik). W niniejszej pracy założono $PL(d_0) = 39$ dB dla transmisji w paśmie 38 GHz i $\alpha = 2.13$ [9].

- dla przeskoku z wykorzystaniem RIS [10]:

$$PL_{RIS}^{dB}(d_{1\to R}, d_{R\to 2}) = PL(d_0) + 10\beta \cdot \log(M^2(d_{1\to R} + d_{R\to 2})) - g_{bf} \quad (2)$$

gdzie M to liczba meta-powierzchni w RIS, $\beta$ to współczynnik tłumienia ścieżki (wykładnik), a $d_{1\to R}$ i $d_{R\to 2}$ to odpowiednio odległości między nadajnikiem i RIS oraz między RIS i odbiornikiem. Dodatkowo założono, że RIS realizuje odbicie sygnału w sposób aktywny, zapewniając jego wzmocnienie $g_{bf}$, dzięki zastosowaniu formowania wiązki. W pracy założono $M = 3$ i $\beta = \alpha = 2.13$.

Średnią przepustowość wybranego łącza można wyznaczyć ze zmodyfikowanego wzoru Shannona [11]:

$$C_i = \eta \cdot B^{(eff)} \log_2(1 + SNR_i), \quad (3)$$

gdzie $\eta$ oznacza efektywność łącza (udział bitów danych w całkowitej ich liczbie), $B^{(eff)}$ to efektywnie wykorzystane pasmo, a $SNR_i$ to średni stosunek mocy sygnału do mocy szumu (SNR) i-tego przeskoku, wyznaczany jako:

$$SNR_i = \frac{P_i^{(TX)} g_i}{\sigma^2}, \quad (4)$$

oznaczając jako $P_i^{(TX)}$ i $g_i$ odpowiednio moc nadawczą (ograniczoną jako $P_i^{(TX)} \leq P_{max}$) i wzmocnienie kanału dla i-tego przeskoku, a moc szumu jako $\sigma^2$. Wzmocnienie kanału wyznaczane jest na podstawie znajomości tłumienia ścieżki $PL_i^{dB}$ obliczanego ze wzoru (1) lub (2) oraz wzmocnienia anteny nadawczej $G_i^{(Tx)}$ i odbiorczej $G_i^{(Rx)}$ jako $g_i = G_i^{(Tx)} G_i^{(Rx)} 10 \log_{10}(PL_i^{dB})$.

### 2.2. Problem optymalizacyjny

Rozważany problem dotyczy maksymalizacji zasięgu sieci zakładając, że dostępna jest jedna MBS zapewniająca łączność z siecią rdzeniową i maksymalnie $N$ UAV działających jako przekaźniki lub AP. Każdy z dronów może zostać umieszczony w jednej z dostępnych lokalizacji $L_n = (x_n, y_n)$, gdzie $(x_n, y_n)$ to współrzędne w układzie kartezjańskim. Zakładając wymóg minimalnej przepustowości łącza dosyłowego $C_{min}$ problem optymalizacyjny można przedstawić jako:

$$\max_{\{L_1, L_2, ..., L_n\}} |A|, n \leq N, \quad (5)$$

gdzie $A = \{L_k, \forall k: Q_k = 1\}$ to zbiór wszystkich możliwych lokalizacji punktu dostępowego AP, dla których istnieje łącze dosyłowe od stacji MBS. $Q_k$ to wskaźnik dostępności łącza dla lokalizacji $L_k$ zdefiniowany jako:

$$Q_k = \prod_{i=1}^{n} q_i, n \leq N, \quad (6)$$

gdzie $q_i$ oznacza dostępność $i$-tego przeskoku łącza dosyłowego określoną następująco:

$$q_i = \begin{cases} 1 & \text{jeśli } C_i \geq C_{min} \\ 0 & \text{w przeciwnym razie} \end{cases} \quad (7)$$

Wartość $C_i$ dla i-tego przeskoku można w praktyce zastąpić wymaganą wartością $SNR_{min}$.

W przypadku użycia RIS, takie połączenie jest uznawane za pojedynczy przeskok, gdyż nie wymaga zastosowania przetwarzania sygnału w paśmie podstawowym w RIS, a wpływ dwóch ścieżek jest uwzględniany w (2).

## 3. PROPONOWANE ROZWIĄZANIE I WYNIKI SYMULACYJNE

### 3.1. Proponowane rozwiązanie dla RIS

W celu znalezienia optymalnej lokalizacji DRS dla realizacji łącza dosyłowego należy rozważyć dla każdego przeskoku, czy między nadajnikiem i odbiornikiem dostępne jest połączenie LoS o określonej przepustowości. W tym celu można zastosować algorytm opisany w [7], gdzie zastosowano algorytm Lee dla określenia grafu widoczności $G(V, E)$, zawierającego zbiór wszystkich krawędzi budynków ($E$) i narożnikowych wierzchołków ($V$), dla których wyznaczana jest widoczność. Dla otrzymanego zbioru $V_g$ zawierającego krawędzie między widocznymi wierzchołkami zastosowano następnie algorytm Dijkstry w celu wyboru najkrótszej ścieżki. W tej pracy rozwinięto algorytm uwzględniający dostępność nowych ścieżek związanych z użyciem RIS umieszczonych na ścianach wybranych budynków. W tym przypadku, dla algorytmu Lee RIS jest traktowany jako wierzchołek w celu znalezienia widoczności, ale nie jest uwzględniany w określaniu liczby przeskoków. Następnie znajdywana jest najkrótsza ścieżka z uwzględnieniem następującej definicji kosztu przeskoku dla algorytmu Dijkstry:

$$D_j = \begin{cases} PL_{direct}^{dB} + P & \text{dla łącza bezp. i } C_i > C_{min} \\ PL_{RIS}^{dB} + P & \text{dla łącza z RIS i } C_i > C_{min} \\ \infty & \text{w przeciwnym razie} \end{cases} \quad (8)$$

gdzie $P$ jest stałym współczynnikiem kary z każdy kolejny przeskok, związanym z koniecznością utraty energii na przetwarzanie sygnału w przekaźniku.

### 3.2. Badania symulacyjne

W celu zbadania wpływu RIS na dostępność łącza dosyłowego przeprowadzono badania symulacyjne dla systemu w dwóch konfiguracjach: bez dostępności RIS i z dwoma matrycami RIS umieszczonymi w centralnej lokalizacji obszaru (na rynku), jak przedstawiono na rys. 3. Pozostałe parametry badanego systemu dobrano zgodnie z wartościami podanymi w Tabeli 1.

Na rys. 2 przedstawiono wyniki badań dla przypadku bez użycia RIS, pokazujące liczbę wymaganych przeskoków dla wybranych lokalizacji punktu dostępowego AP (liczba przeskoków oznaczona kolorem i kształtem markera). W tym przypadku założono $SNR_{min} = 41$ dB w celu określenia dostępności pojedynczego przeskoku. Wyniki te można porównać z rezultatami otrzymanymi dla przypadku z użyciem $R=2$ RIS, przedstawionymi na rys. 3. Porównanie to wyraźnie wskazuje na większy zasięg systemu korzystającego z RIS. Dodatkowo, punkty dostępne również w konfiguracji tylko z wykorzystaniem dronów, w systemie z RIS można osiągnąć stosując mniejszą liczbę DRS.

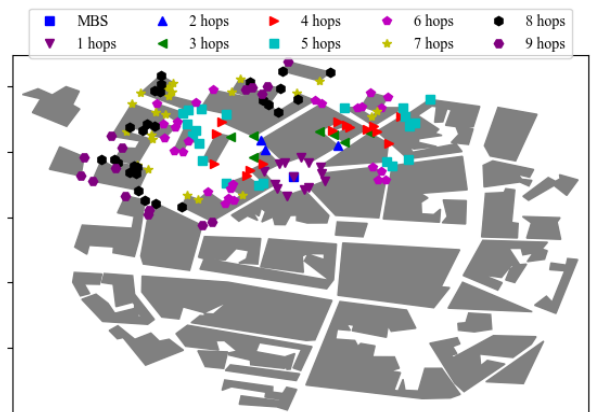

*Rys. 2. Ilustracja niezbędnej liczby przeskoków dla łącza dosyłowego bez zastosowania RIS dla wybranych AP*

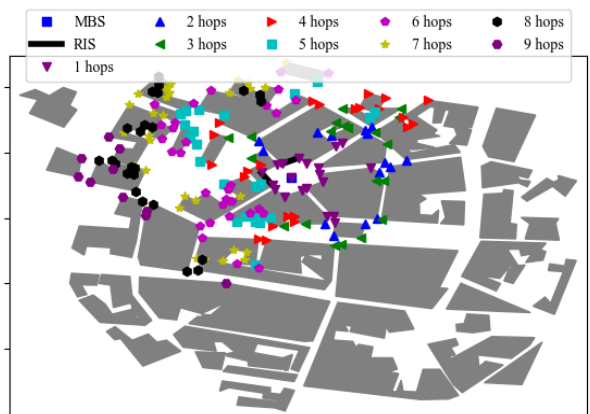

*Rys. 3. Ilustracja niezbędnej liczby przeskoków dla łącza dosyłowego z zastosowaniem RIS dla wybranych AP*

*Tabela 1: Parametry symulacji*

| Parametr | Wartość |
|---|---|
| Moc nadawcza $P^{(Tx)}$ | 100 mW |
| Zysk RIS $g_{bf}$ | 15 dB |
| Moc szumu $\sigma^2$ | -131 dBm |
| Efektywność łącza $\eta$ | 0.82 |
| Efektywne pasmo $B^{(eff)}$ | 18.72 MHz |
| Wymagany SNR: $SNR_{min}$ (odpowiadający $C_{min}$) | 41dB (55 Mbps) |

Kolejne wyniki, przedstawione na rys. 4, pokazują maksymalną możliwą przepustowość łącza dosyłowego w zależności od lokalizacji na mapie korzystając z tzw. mapy cieplnej (ang. *heatmap*). Wartości przepustowości przedstawione na rys. 4 wyznaczono określając najlepszą ścieżkę do wybranego punktu zgodnie z zaproponowaną metodą, a następnie wybierając przepustowość najsłabszego przeskoku łącza dosyłowego. Można zauważyć, że zastosowanie większej liczby RIS w ogólności poprawia przepustowość, jednak w niektórych przypadkach może skutkować jej obniżeniem gdy umożliwia zastosowanie mniejszej liczby DRS. W tym przypadku zysk z zastosowania RIS skutkuje zmniejszeniem zużycia energii.

## 4. PODSUMOWANIE

W niniejszej pracy rozważono problem realizacji bezprzewodowego łącza dosyłowego z wieloma przeskokami między stacją bazową makro i punktem dostępowym, wykorzystując do tego celu drony-przekaźniki i matryce RIS. Zaproponowano problem optymalizacyjny polegający na doborze liczby i lokalizacji dronów w celu zapewnienia odpowiedniej przepustowości łącza, dążąc do zmniejszenia liczby przekaźników lub do maksymalizacji zasięgu sieci. Przeprowadzone badania pokazały, że użycie RIS pozwala wyraźnie zwiększyć zasięg sieci lub zmniejszyć liczbę niezbędnych przekaźników do ustanowienia łącza dosyłowego. Niniejsza praca będzie dalej rozwijana uwzględniając zagadnienia maksymalizacji efektywności energetycznej i minimalizacji opóźnienia transmisji. Dodatkowo pod uwagę można też wziąć możliwość zainstalowania RIS na wybranych dronach zamiast tradycyjnych urządzeń nadawczo-odbiorczych.

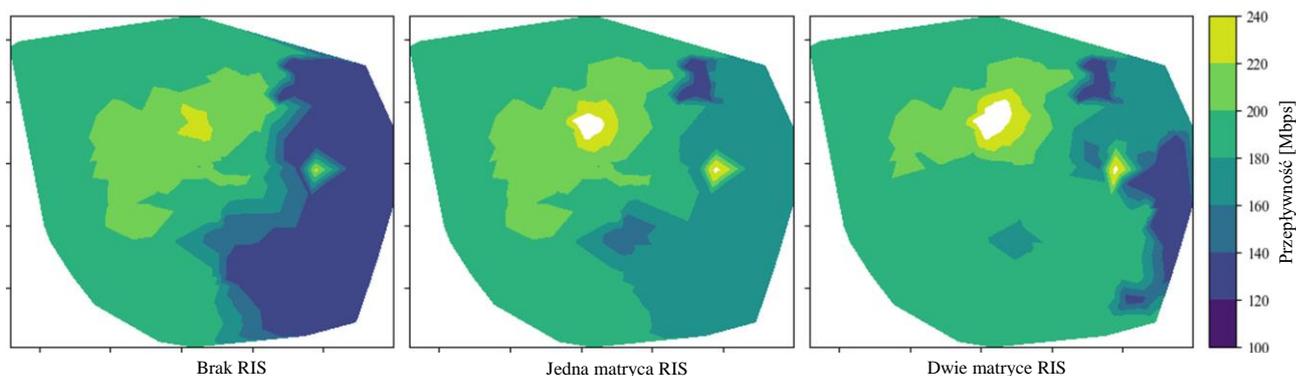

*Rys. 4. Wykresy mapy cieplnej ilustrujące osiągalną przepustowość łącza dosyłowego. Wykres po lewej przedstawia sytuację bez RIS, środkowy - z wykorzystaniem jednej matrycy RIS, a prawy - z zastosowaniem dwóch RIS*